# HealthFog: An Ensemble Deep Learning based Smart Healthcare System for Automatic Diagnosis of Heart Diseases in Integrated IoT and Fog Computing Environments

Shreshth Tuli[1,2], Nipam Basumatary[1,3], Sukhpal Singh Gill[4], Mohsen Kahani[1,5], Rajesh Chand Arya[6], Gurpreet Singh Wander[6], Rajkumar Buyya[1]

## Abstract

Cloud computing provides resources over the Internet and allows a plethora of applications to be deployed to provide services for different industries. The major bottleneck being faced currently in these cloud frameworks is their limited scalability and hence inability to cater to the requirements of centralized Internet of Things (IoT) based compute environments. The main reason for this is that latency-sensitive applications like health monitoring and surveillance systems now require computation over large amounts of data (Big Data) transferred to centralized database and from database to cloud data centers which leads to drop in performance of such systems. The new paradigms of fog and edge computing provide innovative solutions by bringing resources closer to the user and provide low latency and energy efficient solutions for data processing compared to cloud domains. Still, the current fog models have many limitations and focus from a limited perspective on either accuracy of results or reduced response time but not both. We proposed a novel framework called HealthFog for integrating ensemble deep learning in Edge computing devices and deployed it for a real-life application of automatic Heart Disease analysis. HealthFog delivers healthcare as a fog service using IoT devices and efficiently manages the data of heart patients, which comes as user requests. Fog-enabled cloud framework, FogBus is used to deploy and test the performance of the proposed model in terms of power consumption, network bandwidth, latency, jitter, accuracy and execution time. HealthFog is configurable to various operation modes which provide the best Quality of Service or prediction accuracy, as required, in diverse fog computation scenarios and for different user requirements.

## 1. Introduction

Fog and Cloud computing paradigms have emerged as a backbone of modern economy and utilize Internet to provide on-demand services to users [1]. Both of these domains have captured significant attention of industries and academia. But because of high time delay, cloud computing is not a good option for applications requiring real-time response. Technological developments like edge computing, fog computing, Internet of Things (IoT), and Big Data have gained importance due to their robustness and ability to provide diverse response characteristics based on target application [2]. These emerging technologies provide storage, computation, and communication to edge devices, which facilitate and enhance mobility, privacy, security, low latency, and network bandwidth so that fog computing can perfectly match latency-sensitive or real-time applications [2, 6, 10, 12, 27, 40, 45, 46, 47, 48]. Now, cloud computing frameworks also extend support to emerging application paradigms such as IoT, Fog computing, Edge, and Big Data through service and infrastructure [3, 4]. Fog computing uses routers, compute nodes and gateways to provide services with minimum possible energy consumption, network latency and response time.

Mutlag et al. [40] explored the challenges of Fog computing in healthcare applications and identified that latency and response time are the most important and difficult to optimize Quality of Service (QoS) parameters in real time fog environments. Healthcare is one of the prominent application areas that requires accurate and real-time results, and people have introduced Fog Computing in this field which leads to a positive progress. With Fog computing, we bring the resources closer to the users thus decreasing the latency and thereby increasing the safety measure. Getting quicker results implies fast actions for critical heart patients. But faster delivery of results is not enough as with such delicate data we can not compromise with the accuracy of the result. One way to obtain high accuracies is by using state-of-the-art analysis softwares typically those that employ deep learning and their variants trained on a large dataset. In the recent years, deep learning [5] has

---

[1]Cloud Computing and Distributed Systems (CLOUDS) Laboratory, School of Computing and Information Systems, The University of Melbourne, Australia

[2]Department of Computer Science and Engineering, Indian Institute of Technology (IIT), Delhi, India

[3]Department of Computer Science and Engineering, Indian Institute of Technology (IIT), Madras, India

[4]School of Electronic Engineering and Computer Science (EECS), Queen Mary University of London, UK

[5]Web Technologies Laboratory, Ferdowsi University Of Mashhad, Iran

[6]Department of Cardiology, Hero Heart Institute, Dayanand Medical College and Hospital, Ludhiana, Punjab, India

E-mail addresses: shreshthtuli@gmail.com (S. Tuli), nipambasumatary1@gmail.com (N. Basumatary), s.s.gill@qmul.ac.uk (S.S. Gill), kahani@um.ac.ir (M. Kahani), drrajesharya@yahoo.com (R.C. Arya), drgswander@yahoo.com (G.S. Wander), rbuyya@unimelb.edu.au (R. Buyya)

seen an exponential growth in the fields ranging from computer vision [6] to speech recognition, but has more recently been proven useful in natural language processing, sequence prediction, and mixed modality data settings. Moreover, ensemble learning [7] is used to get the best of multiple classifiers. One of the ensemble methods is called bagging classifier where the estimator fits trains the base classifier on random subsets of data and then aggregates their individual predictions either by voting or by averaging to get the final prediction. Such estimators help in reducing the variance as compared to a single estimator by introducing randomization into the dataset distribution procedure. Another advancement of deep learning has been to predict and classify healthcare data with extremely high accuracies [5]. However, recent deep learning models for healthcare applications are highly sophisticated and require large number of computational resources both for training and prediction [8]. It also takes large amount of time to train these complex neural networks and analyze data using them. The higher the accuracy required, the more sophisticated the network and higher is the prediction time [9]. This has been a major problem for healthcare and similar IoT applications where it is critical to obtain results in real-time. As computation on the Edge has the great advantage of reducing response time, this gives a new direction of research of integrating complex ensemble deep learning models with Edge Computing such that we obtain high accuracy results in real-time. One of the fundamental aims of this work is to bridge this gap and provide a computing platform that not only provides low latency results by leveraging edge resources but also is able to use deep learning based frameworks to provide highly accurate results. There has been some work to bring computation to the Edge devices, closer to the patient to reduce result delivery time. Some of these works still depend on simulations [10] and have not provided a deploy-able framework. This work also aims to fill this void in healthcare industry.

Usually, detecting heart problems is difficult [49, 50] and many times people do not even get to know that they are in critical condition till they get heart related problems like tachycardia or even stroke. Conventionally symptoms of heart problems are difficult to identify and requires an experienced doctor to observe the patient to ascertain that he/she has a heart problem. This is difficult to do practically due to shortage of doctors as most countries still do not trust computer systems to be able to detect heart problems with the required accuracy and explain-ability [51, 52]. Existing healthcare systems that are deployed on IoT driven Fog or cloud computing frameworks connect pre-configured devices for patient data processing such that the results are delivered to users within the deadline time. Many prior works have tried to use IoT to predict health problems related to heart but are unable to ascertain with the accuracies required by the stringent regulations of medical standardization agencies. In recent past, as deep learning has gained popularity more recent technologies can even surpass doctors in heart disease detection accuracy [53, 54]. This work aims to bring together deep learning and IoT in healthcare industry in hope that it motivates medical standardization agencies to adopt this model providing low latency and high accuracy to mitigate the problem of lack of doctors. There exist very few works that aim to bring together these two paradigms like [19], but none utilize the distributed nature of edge computing to improve accuracy by utilizing ensemble deep learning models. We present more comprehensive comparisons in Sections 2 and 7.9. Moreover, extension of deep learning models to allow ensembling of results is a non trivial extension as it requires careful balance of accuracy improvement and latency increase to provide the most desired service quality. Furthermore, building on previous works like [2, 19, 46], HealthFog provides a novel architecture for healthcare computation integrating/harnessing diverse backend frameworks like FogBus [27] and Aneka [28] making it a scalable model.

Prior works have reported that there are two major types of healthcare data collection schemes for heart patients using different devices (IoT sensors and file input data). The first is Little data which is processed at fog nodes and the second is Big data processed at Cloud Data Centers (CDC) [1, 3]. The healthcare patient data is received by the network at high speeds (250 MB per minute or more) [1]. Existing frameworks are not versatile enough to capture and provide results for both types of data scenarios and thus there is a need to utilize edge and cloud resources in order to cater to applications with these types of data volumes. Data is stored and processed on edge nodes or cloud servers after collection and aggregation of data from smart devices of IoT networks.

To provide efficient compute services to heart patients and other users requiring real-time results, an integrated Edge-Fog-Cloud based computation model is required to deliver healthcare and other latency sensitive results with low response time, minimum energy consumption and high accuracy. The lack of such models or frameworks that integrate the power of high accuracy of deep learning models simultaneously with low latency of edge computing nodes motivated this work.

In this work, we propose a Fog based Smart Healthcare System for Automatic Diagnosis of Heart Diseases using deep learning and IoT called **HealthFog**. HealthFog provides healthcare as a lightweight fog service and efficiently manages the data of heart patients which is coming from different IoT devices. HealthFog provides this service by using the FogBus framework [27] and demonstrates application enablement and engineering simplicity for leveraging fog resources to achieve the same.

The key **contributions** of this paper are:

- Proposed a generic system architecture for development of ensemble deep learning on fog computing

- Developed a lightweight automatic heart patient data diagnosis system using ensemble deep learning called HealthFog.

- Deployed HealthFog using FogBus framework for integration of IoT-Edge-Cloud for real-time data analysis.

- Demonstrated and analyzed the HealthFog deployment in terms of various performance metrics like accuracy, response time, network bandwidth and energy consumption.

All analysis has been done for heart patient data for prediction if the patient has a heart problem or not.

The rest of the paper is organized as follows. *Section 2* presents related work of existing healthcare systems. Background of FogBus and Aneka are is provided in *Section 3*. Proposed model is presented in *Section 4* and its design and implementation is described in *Section 5*. *Section 7* describes the experimental setup and presents the results of performance evaluation. *Section 8* presents conclusions with future work proposed.

## 2. Related Work

Fog computing environment is an emerging paradigm for efficient processing of healthcare data, which is coming from different IoT devices. Fog computing is capable to handle the data of heart patients at edge devices or fog nodes with large computing capacity to reduce latency, response time or delay because edge devices are closer to the IoT devices than cloud data center.

Gia et al. [11] proposed a Low Cost Health Monitoring (LCHM) model to gather the health information of different heart patients. Moreover, sensor nodes monitor and analyse the Electro Cardio Graphy (ECG) in a real-time manner for processing of heart patients data efficiently, but LCHM has more response time which reduces the performance. Further, sensor nodes gather ECG, respiration rate, and body temperature and transmits to a smart gateway using wireless communication mode to take automatic decision quickly to help patient. Orange Pi One based small-scale testbed is used to test the performance of LCHM model in terms of execution time, but LCHM consumes more energy during collection and transmission of data. He at al. [12] proposed an IoT based healthcare management model called FogCepCare to integrate cloud layer with sensor layer to find out the health status of heart patients and reduces the execution time of job processing at runtime. FogCepCare uses the partitioning and clustering approach and a communication and parallel processing policy to optimize the execution time. The performance of FogCepCare is compared with existing model using simulated cloud environment and optimizes the execution time but this work lacks the evaluation of performance in terms of important QoS parameters such as power consumption, latency, accuracy etc. Ali and Ghazal [13] proposed an IoT e-health service based an application using Software Defined Network (SDN), which collects data through smartphone in the form of voice control and finds the health status of patients. Further, an IoT e-health service finds the type of heart attack using mobile application based conceptual model but performance of the proposed application is not evaluated on cloud environments. Akrivopoulos et al. [14] proposed an ECG-based Healthcare (ECGH) system to diagnose cardiac abnormalities [15] using ECG but has low accuracy and high response time of detecting abnormal events because they are fetching data directly without using data analytics or other feature extraction techniques. Further, the data transmission to cloud server in case of large number of requests increases latency and consumes more energy consumption, which degrades the performance of the system. Manikandan et al. [16] proposed an Autonomous Monitoring System (AMS) model for Internet of Medical Things (IoMT) to provide healthcare facilities. In this research work, a reward-based mechanism designed which utilizes the Analytical Hierarchy Process (AHP) for fair distribution of energy among the nodes. The simulated cloud environment is used to test the performance of the AMS model in terms of energy consumption and AMS model performs better than FGCS method but the communication time among nodes leads to high latency of processing a patient request.

Choi et al. [17] proposed a Graph-based Attention Model (GRAM) for healthcare representation learning that supplements electronic health records with hierarchical information inherent to medical ontologies. Further, the performance of GRAM is optimized in terms of training accuracy. GRAM uses predictive analysis to predict the chances of heart attack and compared the performance of GRAM with Recurrent Neural Network (RNN) using very small dataset and performs better than RNN in terms of training accuracy. The performance of GRAM can be degraded in case of large datasets. Nicholas et al. [18] proposed a Smart Fog Gateway (SFG) model for end-to-end analytics in wearable IoT devices and demonstrated the role of the SFG in orchestrating the process of data conditioning, intelligent filtering, smart analytics, and selective transfer to the cloud for long-term storage and temporal variability monitoring. SFG model optimizes the performance in terms of execution time and energy consumption, but it does not consider latency as a performance parameter. Iman et al. [19] proposed Hierarchical Edge-based deep learning (HEDL) based healthcare IoT system to investigate the feasibility of deploying the Convolutional Neural Network (CNN) based classification model as an example of deep learning methods. Further, a case study of ECG classifications is used to test the performance of proposed system in terms of accuracy and execution time. Liangzhi et al. [20] proposed Fog based Efficient Manufacture Inspection (FEMI) system using deep learning for smart industry to process a large amount of data in an efficient manner. Further, FEMI system adapts the CNN model to the fog computing environment, which significantly improves its computing efficiency and optimizes the performance only in terms of testing accuracy.

Mahmud et al. [21] proposed a Fog-based IoT-Healthcare (FIH) solution structure and explore the integration of Cloud-Fog services in interoperable Healthcare solutions extended upon the traditional Cloud-based structure. Further, iFogSim simulator [43] is used to test the performance of FIH solution in terms of power consumption and latency only. The performance of FIH solution can be evaluated in terms of execution time and accuracy. Rabindra and Rojalina [22] proposed a fog-based machine learning model for smart system big data analytics called FogLearn for application of K-means clustering in Ganga River Basin Management and real-world feature data for detecting diabetes patients suffering from diabetes mellitus. Alvin et al. [23] proposed a Scalable and Accurate deep learning (SADL) model with electronic health records of patients based on the Fast Healthcare Interoperability Resources (FHIR) format. The deep learning methods in SADL model using FHIR representation are capable of accurately predicting multiple medical events from multiple centers without site-specific data harmonization. Further, proposed approach is validated using de-identified Electronic Health Record (EHR) data from two US academic medical centers with 216,221 adult patients hospitalized for at least 24 hours and improves the accuracy of prediction. Table 1 compares the proposed model (HealthFog) with existing models.

| Work | Fog Computing | IoT | Deep Learning | Ensemble Learning | Heart Disease Prediction System | Power Consumption | Latency | Performance Parameters | | | Jitter | Testing Accuracy | Training Accuracy |
|---|---|---|---|---|---|---|---|---|---|---|---|---|---|
| | | | | | | | | Execution Time | Arbitration Time | Network Bandwidth | | | |
| LCHM [11] | ✓ | | | | | | | ✓ | | | | | |
| FogCepCare [12] | ✓ | | | | | | | ✓ | | | | | |
| IoT e-health service[13] | | ✓ | | | ✓ | | | | | | | | |
| ECGH [14] | ✓ | ✓ | | | | | | | | | | ✓ | |
| AMS [16] | | ✓ | | | | ✓ | | | | | | | |
| GRAM [17] | | | ✓ | | ✓ | | | | | | | | ✓ |
| SFG [18] | ✓ | ✓ | | | | ✓ | | ✓ | | | | | |
| HEDL [19] | ✓ | ✓ | ✓ | | | | | ✓ | | | | | ✓ |
| FEMI [20] | ✓ | | ✓ | | | | | | | | | ✓ | |
| FIH [21] | ✓ | ✓ | | | | ✓ | ✓ | | | | | | |
| FogLearn [22] | | | ✓ | | | | | | | | | | |
| SADL [23] | | | ✓ | | | | | | | | | ✓ | |
| CoSHE [39] | | ✓ | | | | | | | | | | | |
| EOTC [41] | ✓ | ✓ | | | | | | | | | | | |
| SLA-HBDA [42] | | | | | | | | ✓ | | | | ✓ | |
| CFBA [43] | ✓ | | | | | | ✓ | | | | | | |
| **HealthFog (this work)** | ✓ | ✓ | ✓ | ✓ | ✓ | ✓ | ✓ | ✓ | ✓ | ✓ | ✓ | ✓ | ✓ |

Table 1: Comparison of existing models with HealthFog

Pham et al. [39] proposed a Cloud-based Smart Home Environment (CoSHE) to deliver home healthcare to provide humans contextual information and monitors the vital signs using robot assistant. Initially, CoSHE uses non-invasive wearable sensors to gather the audio, motion and physiological signals and delivers the contextual information in terms of the residents daily activity. Further, the CoSHE allows healthcare professionals to explore behavioural changes and daily activities of a patient to monitor the health status periodically. Moreover, the case study of robotic assistance is presented to test the performance of CoSHE by utilizing Google APIs. However, CoSHE is general healthcare application to collect and process patient data at small scale without data analytics and they have not evaluated on real cloud environment to test its performance in terms of QoS parameters.

Alam et al. [41] proposed a general Edge-of-Things Computation (EoTC) framework for healthcare service provisioning to optimize the cost of data processing. Further, a portfolio optimization solution is presented for the selection of Virtual Machines (VMs) and designed Alternating Direction Method of Multipliers (ADMM) based distributed provisioning technique for efficient processing of healthcare data. Further, experimental results demonstrate that EoTC framework performs better than greedy approach in terms of cost, but this framework lacks in performance evaluation in terms of QoS parameters.

Sahoo et al. [42] proposed a Service Level Agreement (SLA) based Healthcare Big Data Analytic (SLA-HBDA) architecture to perform the ranking of patients data, which improves its pro-cessing speed. Further, an efficient data distribution technique is developed to allocate batch and streaming data using Spark platform to predict the health status of the patient. SLA-HBDA architecture improves the performance in terms of accuracy as compared to Naive-Bayes (NB) algorithm but it does not consider latency and other important QoS parameters.

Abdelmoneem et al. [43] proposed a Cloud-Fog Based Architecture (CFBA) for IoT based healthcare applications to monitor the health status of the patience. Further, a task scheduling and allocation mechanism is proposed for the processing of healthcare data by distributing the healthcare tasks in an efficient manner. The performance of CBFA is evaluated using iFogSim simulator [44] in terms of only latency. Research work [39, 41, 42, 43] developed general healthcare applications at small scale and none of the work focused on heart patient-based healthcare application to diagnose the health status of heart patients.

Sanaz et al. [46] proposed an end-to-end security scheme for mobility enabled healthcare IoT, which uses Datagram Transport Layer Security (DTLS) handshake protocol to establish secure communication among various interconnected smart gateways without requiring any reconfiguration at the device layer. Further, the proposed scheme is implemented using simulation environment (Cooja) and demonstrate that the proposed scheme is effective in reducing communication overhead by 26% and latency by 16%. Building on this work, HealthFog aims to deploy healthcare applications on real systems and fog nodes providing a more promising solution.

Amir et al. [2] proposed a system called Smart e-Health Gateway to exploit the strategic position of such gateways at the edge of the network to provide various services such as embedded data mining, real-time local data processing and local storage. Further, it distributes the burden of various sensors by creating a Geo-distributed intermediary layer of intelligence between Cloud and sensor nodes, which increases the reliability, energy efficient and scalability. Further, proposed system is validated using an mobile application of IoT-based Early Warning Score (EWS) health monitoring. Building on this work, Health-Fog architecture provides additional features of being able to use distributed deep learning models in ensembling fashion to further increase the prediction accuracy and provide more precise results for critical heart patients.

There is a need to solve the following challenges [24, 25, 11, 12, 13, 14, 16, 17, 18, 19, 20, 21, 22, 23, 26, 39, 42, 43, 44] to recognize the full capability of IoT based fog-computing for healthcare systems: (a) An efficient IoT based Healthcare application is needed which can process a large amount of heart patients data with minimum energy consumption and low response time, (b) a well-organized resource scheduling technique is needed for fog computing environments to execute user workloads with maximum resource utilization to fulfill the deadline of workloads and (c) ensemble deep learning based fog computing model to automatically diagnose the heart disease severity in patients in real-time.

## 3. Background Technologies

FogBus [27] is a framework for development and deployment of integrated Fog-Cloud environments with structured communication and platform independent execution of applications. FogBus connects various IoT sensors which can be healthcare sensors with gateway devices to send data and tasks to fog worker nodes. The resource management and task initiation is done on fog broker nodes. To ensure data integrity, privacy and security, FogBus uses blockchain, authentication and encryption techniques which increase the reliability and robustness of the fog environment. FogBus uses HTTP RESTful APIs for communication and seamlessly integrates fog setup with Cloud using Aneka software platform [28].

Aneka [28] is a software platform and framework facilitating the development and deployment of distributed applications onto clouds. Aneka provides developers with APIs for exploiting virtual resources on the cloud. The core components of the Aneka framework are designed and implemented in a service-oriented fashion. Dynamic provisioning is the ability to dynamically acquire resources and integrate them into existing infrastructures and software systems. In the most common case, resources are Virtual Machines (VMs) acquired from an Infrastructure-as-a-Service (IaaS) cloud provider. Dynamic provisioning in Aneka happens as part of the Fabric Services by offering provisioning services for allocating virtual nodes from public cloud providers to complement local resources. This is mainly achieved as a result of the interaction between two services: the Scheduling Service and the Resource Provisioning Service. Aneka currently supports four different programming models [28]: Bag of tasks model, Distributed threads model, MapReduce model, and Parameter sweep model. In Health-Fog, we used the Bag of tasks model for task distribution across cloud VMs. HealthFog uses FogBus to harness fog resources and Aneka to harness cloud resources.

## 4. System Architecture

The HealthFog model is an IoT based fog-enabled cloud computing model for healthcare, which can manage the data of heart patients effectively and diagnose the health status to identify heart disease severity. HealthFog integrates diverse hardware instruments through software components and allows structured and seamless end-to-end integration of Edge-Fog-Cloud for fast and accurate delivery of results. Figure 1 presents the architecture of HealthFog which comprises of various hardware and software components that are described next.

### 4.1. HealthFog hardware components

The HealthFog model comprises of following hardware components:

1. **Body Area Sensor Network**: Three different types of sensors constitute this component: medical sensors, activity sensors and environment sensors. Medical sensors include Electro Cardio Gram (ECG) sensor, Electro Encephalo Gram (EEG) sensor, Electro Myo Graphy (EMG) sensor, oxygen level sensor, temperature sensor, respiration rate sensor and glucose level sensor. This component senses the data from heart patient and transfers to connected gateway devices.

2. **Gateway**: There are three different types of Gateway devices (mobile phones, laptop and tablets), which are acting as a fog device to collect sensed data from different sensors and forward this data to Broker/Worker nodes for further processing.

3. **FogBus Modules**: The FogBus framework comprises of the following:

   (a) **Broker node**: This component receives the job requests and/or input data from Gateway devices. Request input module receives job requests from Gateway devices just before transferring the data. Security Management module provides secure communication between different components and protects the collected data from unauthorized access or malicious tampering of data to improve system credibility and data integrity. Arbitration module (part of Resource Manager in broker node) takes as input the load statistics of all worker nodes and decides which node or subset of nodes to send jobs to in real time.

   (b) **Worker node**: This is the component that performs tasks allocated by the Resource Manager of the Broker node. Worker nodes can comprise of embedded devices and Single Board Computers (SBC) like Raspberry Pis. In HealthFog, Worker nodes can contain sophisticated deep learning models to process and analyse the input data and generate results. Apart from this, the Worker node can include other components for data processing, data filtering and mining, Big Data analytics and storage. The Worker nodes directly get the input data from the Gateway devices, generate results and share with the same. In Health-Fog model, the Broker node can also behave as a Worker node.

   (c) **Cloud Data Center**: When the fog infrastructure becomes overloaded, services are latency tolerant or the input data size is much larger than average size, then

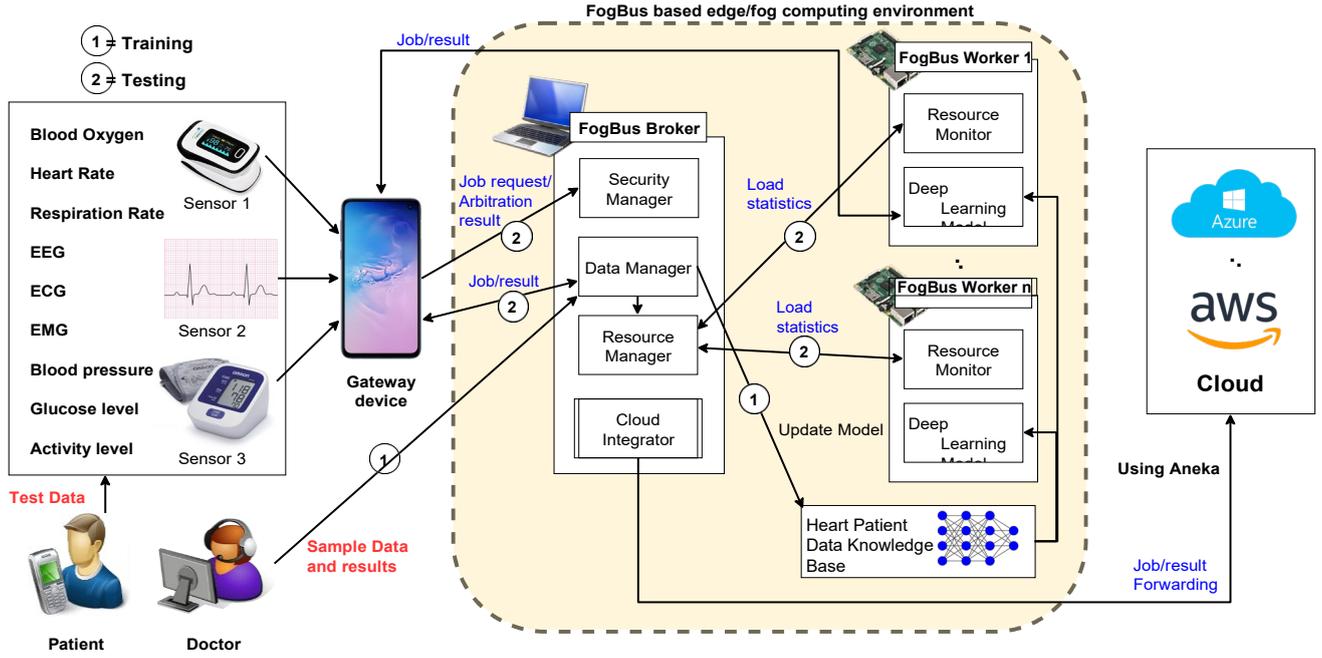

Figure 1: HealthFog Architecture

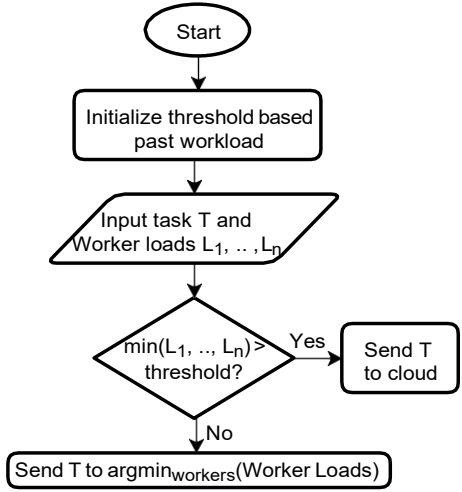

Figure 2: Resource Scheduling in HealthFog

HealthFog harnesses resources of Cloud Data Centers (CDC). This makes it more robust, capable of performing heavy load tasks quickly and also makes data processing location independent.

### 4.2. HealthFog software components

The HealthFog model comprises of the following software components:

- **Data filtering and pre-processing**: The first step after data input is to pre-process it. This includes data filtering using data analytics tools. The filtered data is reduced to a smaller dimension using Principal Component Analysis (PCA) using Set Partitioning In Hierarchical

Trees (SPIHT) algorithm [29] and encrypted using Singular Value Decomposition [30] technique with the goal of extracting key components of data feature vectors that affect the health status of patients. Based on the extracted data, it automatically makes the decision, which recommends medication and suitable check-up based on the continuous training data of healthcare providers and doctors and stores in database for re-training when required.

- **Resource Manager**: This comprises of two modules: workload manager and arbitration module [27]. Workload manager maintains job request and task queues for data processing. It also handles bulk of data which needs to be processed. The Arbitration module schedules the provisioned fog or cloud resources for processing of tasks queued and maintained by the workload manager. Arbitration module resides in the Broker node and decides which Fog computing node should be forwarded the data to obtain the results, the Broker itself, Fog worker node or the Cloud Data Center [27]. The main goal is to divide tasks to different devices to balance load and provide optimum performance. HealthFog allows users to set their own load balancing and arbitration schemes based on the application requirements. The current scheme is described as a flowchart in Figure 2.

- **Deep learning Module**: This module uses the dataset to train a Neural Network to classify data-points which are feature vectors obtained after pre-processing the data obtained from the Body Area Sensor Network. Based on the task allocated by the Resource Manager, it also predicts and generates results for the data obtained from the Gate-way devices.

- **Ensembling Module**: This module receives prediction results from different models and uses voting to decide the output class which is whether the patient has heart disease or not. This module resides in the FogBus node which is assigned the task and is responsible for distributing data and collecting results from other worker nodes.

### 4.3. HealthFog topology

The HealthFog components described previously share large amount of data, information and control signals among themselves. To facilitate this stable network communication is necessary. In addition, the communication should be persistent and fault-tolerant. Taking all these into account, the components are structured in a topology shown in Figure 1. The communication across all devices on the Edge is facilitated using FogBus [27] and that with Cloud VM is using Aneka [28].

The Network topology in HealthFog follows Master-Slave fashion where the Broker Node (Master) controls the Worker Nodes (Slaves). In HealthFog all the edge devices including the Gateway devices, Broker node and Worker nodes are present in the same Local Area Network (LAN). The Resource Manager software component resides in the Broker Node and thus the Gateway devices send job requests to it. The arbitration results obtained from the Resource Manager is received by the Gateway device which instructs it where to send the data. Three scenarios arise here: (1) Broker processing data as Worker Node, (2) Another Worker node to send data and (3) Cloud Data Center based processing. Based on the scenario, the Gateway device may send the data directly to Worker node or Broker node (with/without cloud forwarding). Broker may provide computation services for tasks only when it has sufficient resources and/or the worker nodes are overloaded. If the data is to be forwarded to Cloud, then it goes through the Broker node as the Gateway may not have access to the Virtual Private Network (VPN) in which the Cloud Virtual Machine is present. Apart from this, the Worker nodes periodically send heartbeat packets to the Broker to indicate that they are alive. These packets also include load information that is used by the Resource manager for load balancing.

### 4.4. Sequence of communication

In HealthFog, all hardware components interact based on predefined protocols described in Figure 3 for the three scenarios defined earlier: Broker Only, Worker Node or Cloud. In every scenario the Gateway first sends a Job request to the Broker node. Based on the scenario, the Broker node sends the Gateway either the Worker IP address (of the same LAN) or Master IP address (with/without cloud forwarding). In the Broker only case, the Broker node may or may not check loads of workers. If all workers have heavy loads or all are compromised and Cloud is disabled, then the Broker sends the Gateway devices its IP without cloud forwarding. If there exist workers not heavily loaded then the Broker sends the IP address of least loaded Worker node to the Gateway device. Increasing the number of Workers would increase the arbitration time as more load checks need to be done. In non-cloud case, the Gateway device sends job i.e. input data for analysis to Worker/Broker node which then run pre-processing, prediction model and send results back to Gateway device. In cloud forwarding case, as the Gateway device may not be on the VPN, so it sends the input data to Broker node which then forwards it to the CDC. This also ensures that the IoT sensors and gateway devices are protected from malicious entities and hackers as they may not be connected to Internet but only the LAN with other Fog nodes. Due to larger resource availability at Cloud, the Execution time is expected to be lower but latency higher due to communication overheads and queuing delay at both Broker and CDC. When ensemble is enabled then the data received by the Broker/worker node is forwarded to all other edge nodes and majority class is chosen by the worker node to which the data was sent using bagging.

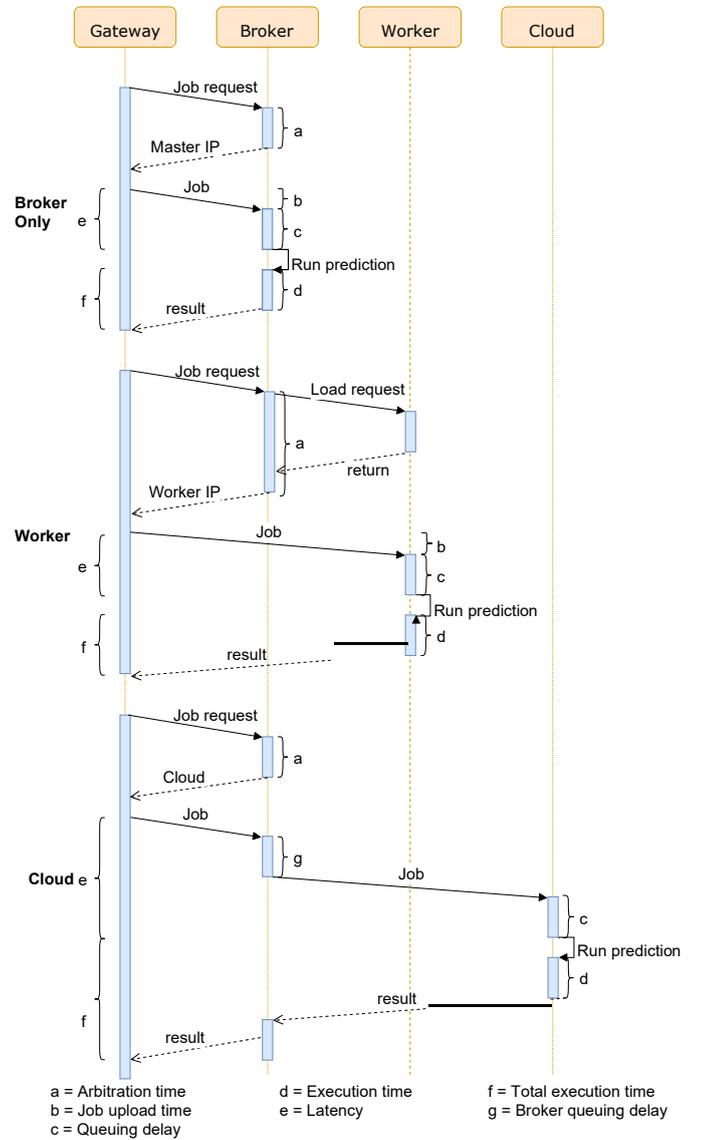

Figure 3: Communication sequence in HealthFog

## 5. HealthFog Design

The fog computing model described in *Section 4* takes heart patient data as input from the sensors and sends back results which comprise of whether the patient has heart disease or not, with the confidence of the claim. This is implemented with components which include data pre-processing modules, ensemble deep learning modules and gateway interface described next.

### 5.1. Heart Patient Data pre-processing

The data obtained from common pule-oximeters or ECG devices is in plain graphical format and needs to be pre-processes to find values of many features of the input to the deep learning model [31, 32]. This requires application specific domain knowledge to be fed into the system. Normalising the age data as it was slightly skewed as shown in Figure 4. Similarly, the Rest Blood Pressure (BPS) data is also skewed and patients having a heat disease showed a higher blood pressure compared to patients not having a heart disease. Patient cholesterol levels also show some target specific behavior, the healthy patients distribution is leptokurtic. Even with maximum heart rate, healthy people have quite higher maximum heart rate (around 160) compared to those with heart disease (around 150). Other features like chest pain and fasting blood sugar had to be converted from continuous values to categorical values. Also, the slope of the peak exercise ST segment and the heart status as retrieved from Thallium test.

### 5.2. Ensemble Deep learning Application

We have used an ensemble of deep neural network as a model for the predictive analysis, and for our application the model is used for binary classification problem. The model is first trained on the heart patient data in the Cleveland Dataset and corresponding known output class and then the trained model is used for predicting results of real time data input as shown in Figure 5.

We divide the data into training, validation and testing set in the ratio of 70:10:20. The training set is used for training the model, the validation set is used for tuning the model and the test set is used for testing how the model performs on new data. The trained model can be stored in all the nodes which are capable of processing by first storing in a common database. Other approach can be to train models separately by distributing the training dataset points across different models. In distributed training, data distribution uses techniques like *boosting* which randomly samples data from the dataset with replacement and sends to different edge nodes for training individual models [7]. At diagnosis time, whenever a node is assigned a task, it gets the patients data which is a vector of size 13. This data is fed as input to the model, makes a forward pass on the deep neural network and outputs 1 or 0 i.e whether the patient has heart disease or not. At diagnosis time, we use the ensemble method of Bagging to combine the results of various models to provide more accurate results. The worker that gets the input data multicasts it to other worker nodes. Each worker then adds this to its queue and the prediction results of each worker node are sent back to the worker assigned for this task. Then the majority prediction class obtained in by bagging is sent it to the gateway device. HealthFog allows users to disable this feature when the results needed are latency critical. In *Section 7* we show that ensemble learning gives better accuracies but also has higher response time and network overheads.

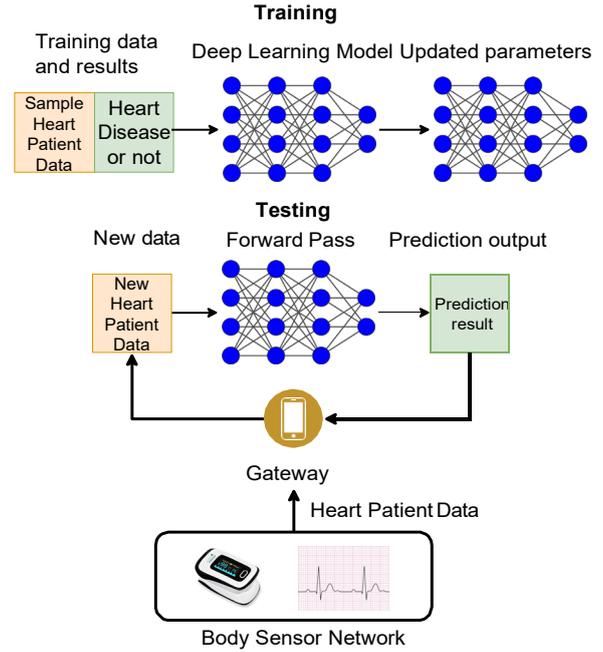

Figure 5: Training and Testing of the application

### 5.3. Android interface and Communication

An android executable named *FastHeartTest* was used in the Gateway device to send data to the Broker/Worker nodes. The application interface is shown in Figure 6. This application allows the Gateway to act as a mediator between the Body Sensor Network and the Worker nodes. The communication is achieved using HTTP RESTful APIs. We used HTTP POST to upload input data from and download results to the Gateway device. Each Worker node, the Broker node and CDC contains a pre-trained deep learning model and pre-processing softwares.

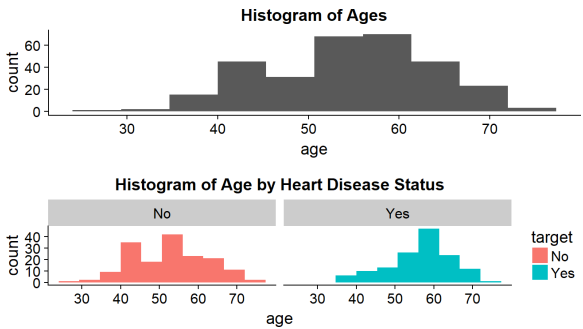

Figure 4: Age distribution

## Gateway Interface screenshots

**(a) Negative example** — Fast Heart Testing form with values: Master IP 192.168.43.130, Age 63, Sex 1, Chest Pain 3, Rest BPS 145, Cholestrol 233, Fasting blood sugar 1, Thalach 150, Exercise induduced angina 0, Oldpeak 2.3, Slope 0, Flouropsy 0, Thal 1. Work sent to 192.168.43.130. Result: You are safe, no need to worry.

**(b) Positive example** — Fast Heart Testing form with values: Master IP 192.168.43.130, Age 54, Sex 1, Chest Pain 2, Rest BPS 150, Cholestrol 232, Fasting blood sugar 0, Thalach 165, Exercise induduced angina 0, Oldpeak 1.6, Slope 2, Flouropsy 0, Thal 1. Work sent to 192.168.43.130. Result: You have heart disease, please consult doctor.

Figure 6: Gateway Interface of HealthFog

## 6. Implementation

The components mentioned in Section 5 were implemented in various programming languages. The pre-processing and ensemble deep learning components were implemented using Python. The pre-processing module normalizes the data based on the maximum and minimum values of the field parameters in the dataset and their distribution.

The ensemble deep learning application used SciKit learn Library [33]. We have used *BaggingClassifier* of the SciKit learn Library to implement our voting scheme. The model takes the type of base classifier which is deep neural network in our case and the number of classifiers as input. Now the model randomly distributes the data among the classifiers to train them. At diagnosis time it takes all predicted classes as input and outputs the majority prediction. The following are the parameters of the best base model on our data set after tuning:

- Size of input layer: 13 (number of features of the data)

- Size of output layer: 2 (Binary classification; whether the patient has heart disease or not)

- Number of hidden layers: 3

- Layer descriptions: Fully connected (FC) layer with 20 nodes, FC layer with 20 nodes and FC Layer with 10 nodes

- Optimizer: Adam

- Activation function: ReLU

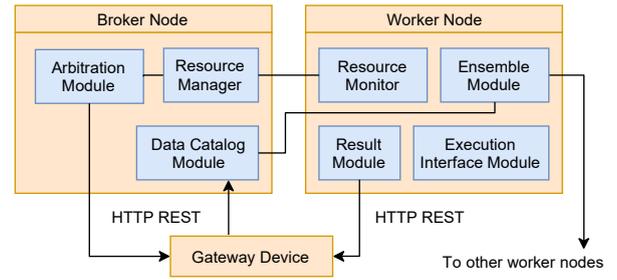

Figure 7: Different modules in HealthFog

- Learning rate: 0.0001

The Android application was built using MIT's App Inventor[1] and communicated with the FogBus Broker node. The android application saves the data attributes in a Comma Separated Value (.csv) file and uploads it to the broker node using HTTP POST to the Data Catalogue Module.

The broker node also has an Arbitration Module which decides which worker node to select for task execution. This worker selection process is done as per the default FogBus policy of selecting worker with minimum CPU load. Whichever worker is selected, is sent the CSV file for analysis. The Execution Interface Module in each worker receives the data and instantiates the Ensemble Deep Learning code for analysis of the data. The returned result is sent back to the Worker/Broker node which sent the data file. The result is ensembled using the bagging strategy and forwarded to the gateway device (android application).

A diagrammatic representation of different modules and their interaction is shown in Figure 7.

## 7. Performance Evaluation

To demonstrate the feasibility and efficacy of the proposed HealthFog model, we implemented and deployed it on actual Fog framework of devices using the FogBus framework [27]. The model has been used for a real-world application of detecting Heart problems for patients instantly using state-of-the art deep learning techniques using a Fog based computing environment. We have analyzed the accuracy and response times with network and energy overheads to show that the HealthFog model is productive and has low overheads.

### 7.1. Experimental Setup

The system setup for the HealthFog evaluation and the hardware configurations are described below:

- **Gateway Device**: Samsung Galaxy S7 with android 9

- **Broker/Master Node**: Dell XPS 13 with Intel(R) Core(TM) i5-7200 CPU @ 2.50GHZ, 8.00 GB DDR4 RAM and 64-bit Windows 10. The deployment used Apache HTTP Server 2.4.34.

---

[1]MIT App Inventor 2: http://ai2.appinventor.mit.edu/

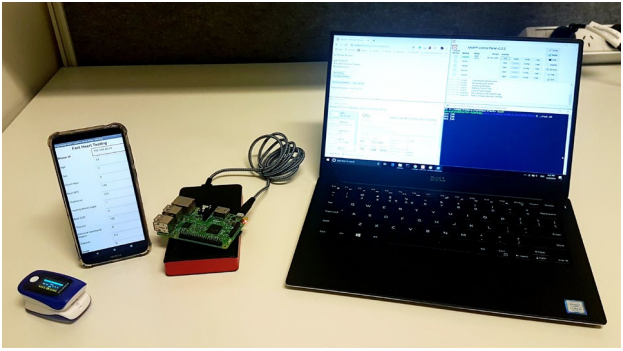

Figure 8: Real HealthFog deployed model and test setup

- **Worker Node**: Raspberry Pi 3B+, ARM Cortex-A53 quad-core SoC CPU @ 1.4 GHz and 1GB LPDDR2 SDRAM and IEEE 802.11 Wifi. Raspbian Stretch Operating system with Apache HTTP server 2.4.34.

- **Public Cloud**: Microsoft Azure B1s Machine, 1vCPU, 1GB RAM, 2GB SSD, Windows Server 2016.

Figure 8 depicts the real implementation of this system model. During the experiments, data parameters are recorded using Microsoft Performance Monitor at the Master and the Azure VM whereas at the Raspberry Pi circuits NMON Performance Monitor is used [34, 35]. To measure the network bandwidth consumption Microsoft Network Monitor 3.4 was used at the Broker node [36] and the vnStat [37] tool in Raspberry Pis.

### 7.2. Dataset

For the experimental results, we have considered the data of heart patients to find the presence of heart disease in the patient [26, 38, 31, 32], which is an integer valued 0 (no presence) or 1 (presence). The Cleveland database [26] is used to conduct the experiments which was created by Andras Janosi (M.D.) at the Gottsegen Hungarian Institute of Cardiology, Hungary and others. The patient names and their patient numbers are kept confidential. We have used 14 important attributes of data to find out the status of patient health: (1) age: age in years, (2) sex: two values (1 = male; 0 = female), (3) cp: chest pain type: - Value 1: typical angina – Value 2: atypical angina – Value 3: non-anginal pain – Value 4: asymptomatic, (4) trestbps: resting blood pressure (in mm Hg on admission to the hospital), (5) chol: serum cholesterol in mg/dl, (6) fbs: (fasting blood sugar > 120 mg/dl) (1 = true; 0 = false), (7) restecg: resting electro-cardiographic results – Value 0: normal – Value 1: having ST-T wave abnormality (T wave inversions and/or ST elevation or depression of > 0.05 mV) – Value 2: showing probable or definite left ventricular hypertrophy by Estes' criteria, (8) thalach: maximum heart rate achieved, (9) exang: exercise induced angina (1 = yes; 0 = no), (10) oldpeak = ST depression induced by exercise relative to rest, (11) slope: the slope of the peak exercise ST segment – Value 1: upsloping – Value 2: flat – Value 3: downsloping, (12) ca: number of major vessels (0-3) colored by flourosopy, (13) thal: 3 = normal; 6 = fixed defect; 7 = reversable defect, (14) target (num): diagnosis of heart disease (angiographic disease status) – Value 0: < 50% diameter narrowing – Value 1: > 50% diameter narrowing (in any major vessel). Table 2 describes the details of just 10 heart patients.

### 7.3. Framework characteristics experiments

Using the dataset mentioned in Section 7.2, we test our model on how well it performs to predict if the patient has a heart disease or not based on the values of the parameters specified for each patient. The dataset was divided into two parts of 70%, 10% and 20% of the whole data. The first part was used to train the model, the second for validation and tweaking the model parameters. The last part was used for testing the model performance. To measure the performance of the HealthFog model the following characteristics were observed and analyzed:

1. **Prediction accuracies**: The dataset consists of 1807 examples out of which 1355 were used for training the model and 452 were used for testing. The training examples were divided equally across all worker/broker nodes equally to obtain their respective trained deep learning models. As the number of Fog nodes increases to use all resources for training the dataset examples would have to be distributed to all nodes. This reduces the training time but also the test accuracy. To observe such effects, the training and test accuracies were analyzed. We define accuracy more formally as the percentage of the total patients for which the model predicts correctly if they have heart disease or not. We compare accuracies for different fog settings, by changing the number of edge nodes and with or without ensembling of results.

2. **Time characteristics**: A representative subset of the different timing parameters shown in Figure 3 were also observed and studied. These include arbitration time, latency, execution time and jitter. We compare these timing parameters for different fog settings by having no edge nodes or upto 2 edge nodes (with or without ensembling) or having a cloud only computation infrastructure.

3. **Network bandwidth usage**: As the scenario i.e. Broker only, Workers or Cloud and the number of Worker nodes affect the network consumption this was studied to find out the network usage in different cases. Similar to the experiments for timing parameters, we compare the network bandwidth consumption for the different fog scenarios. This was done to find out the dependence of bandwidth consumption with different fog configurations that HealthFog provides.

4. **Power consumption**: Energy being a crucial reason of shift from cloud to fog domains, we also studied the power consumption in different scenarios. Based on the power consumption studies and other experiments described earlier we discuss how different HealthFog configurations can be used for various user and application requirements.

### 7.4. Prediction Accuracies

Figure 9 shows the variation of training accuracy with number of Edge nodes (Broker plus Worker nodes). We can observe that the training accuracy gradually increases as the number of worker nodes increase. This is because each node learns a (probability of no disease) and $p_1$ (probability of heart disease), such that $p_0 + p_1 = 1$. The confidence measure of a prediction $(p_0, p_1)$ is quantified as $100 \times (2 \times max(p_0, p_1) - 1)$ and thus has range [0,100]. Thus, if prediction probabilities is (0.5, 0.5) then the confidence is 0 and when they are (0.9, 0.1) then the prediction class is 0% and confidence is 80%. Figure 11 shows the variation of confidence of the binary classifier for the complete test dataset, subset on which the model predicted correctly and that where prediction was incorrect. We see that the confidence is higher for the datapoints where the prediction was correct compared to those datapoints where the prediction was incorrect. The maximum confidence with which the model predicts incorrectly is 49.7%, thus if confidence is less that 50% then our model suggests the patient to consult the doctor as the prediction may be unreliable.

| age | sex | cp | trestbps | chol | fbs | restecg | thalach | exang | oldpeak | slope | ca | thal | target |
|---|---|---|---|---|---|---|---|---|---|---|---|---|---|
| 63 | 1 | 3 | 145 | 233 | 1 | 0 | 150 | 0 | 2.3 | 0 | 0 | 1 | 1 |
| 37 | 1 | 2 | 130 | 250 | 0 | 1 | 187 | 0 | 3.5 | 0 | 0 | 2 | 1 |
| 41 | 0 | 1 | 130 | 204 | 0 | 0 | 172 | 0 | 1.4 | 2 | 0 | 2 | 1 |
| 56 | 1 | 1 | 120 | 236 | 0 | 1 | 178 | 0 | 0.8 | 2 | 0 | 2 | 1 |
| 57 | 0 | 0 | 120 | 354 | 0 | 1 | 163 | 1 | 0.6 | 2 | 0 | 2 | 1 |
| 62 | 0 | 0 | 140 | 268 | 0 | 0 | 160 | 0 | 3.6 | 0 | 2 | 2 | 0 |
| 63 | 1 | 0 | 130 | 254 | 0 | 0 | 147 | 0 | 1.4 | 1 | 1 | 3 | 0 |
| 53 | 1 | 0 | 140 | 203 | 1 | 0 | 155 | 1 | 3.1 | 0 | 0 | 3 | 0 |
| 56 | 1 | 2 | 130 | 256 | 1 | 0 | 142 | 1 | 0.6 | 1 | 1 | 1 | 0 |
| 48 | 1 | 1 | 110 | 229 | 0 | 1 | 168 | 0 | 1 | 0 | 0 | 3 | 0 |

Table 2: Sample patient record data from Cleveland database

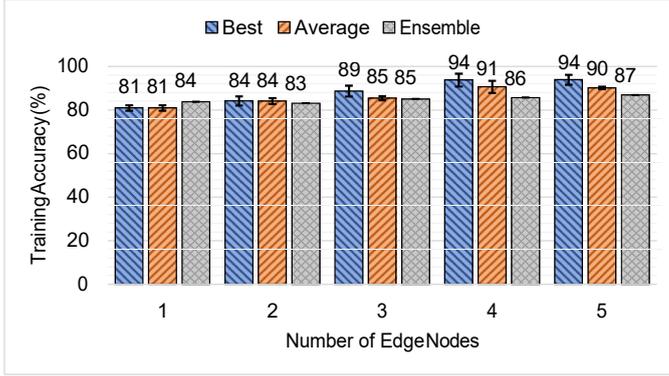

Figure 9: Training accuracy with number of edge nodes

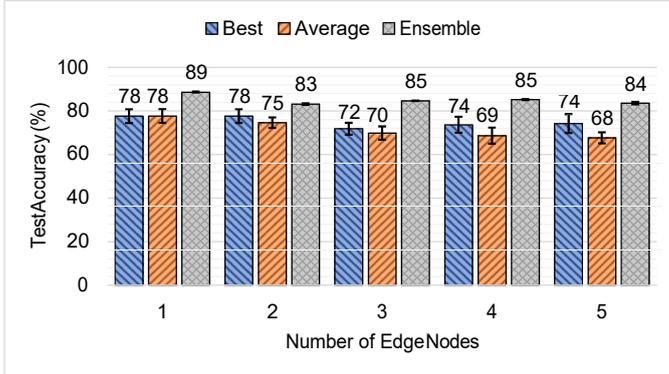

Figure 10: Test accuracy with number of edge nodes

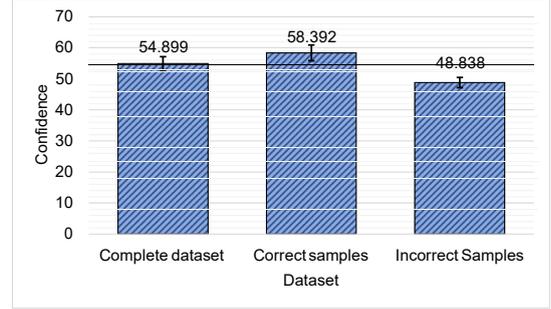

Figure 11: Confidence of the model for different subsets of Cleveland Data

### 7.6. Timing Characteristics

Figure 12 shows the variation of arbitration time at the Broker node for different Fog scenarios: (1) Broker only, (2) Single Worker node, (3) Two worker nodes and (4) Cloud. We see that arbitration time is negligible (nearly 115 ms) when the task is to be sent directly to Broker/Master or Cloud. As the number of edge nodes increase, the Broker needs to check loads at every Worker node and find the minimum load worker to send task, hence the arbitration time increases as number of Edge nodes increase. When the data is sent to worker nodes for ensemble learning, then also the broker does not need to do any load checking as majority class choice needs to be done by one of the worker nodes, thus arbitration time is similar to without ensembling case.

Figure 13 shows the variation of latency, which as per Figure 3 is the addition of communication time and queuing de-model for the data received by it, and as the number of nodes increase, the number of examples received by each node becomes lesser and hence training the model for multiple epochs over-fit the samples and hence training accuracy increases. Figure 10 shows the variation of test data accuracy as the number of Edge nodes increase. As expected, test accuracy decreases with higher number of nodes because each node gets a smaller subset of training data and hence is unable to generalise the model. Another observation is that ensemble learning always gives much better accuracy than the without ensemble case (best or average).

### 7.5. Prediction Confidence

Whenever the deep learning model predicts whether the patient has heart disease or not it generates two probabilities: $p_0$

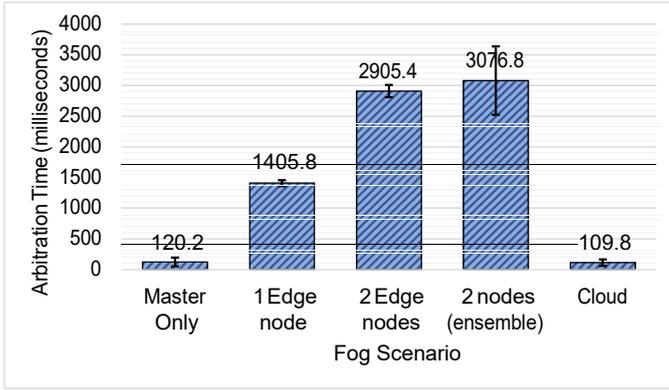

Figure 12: Arbitration time in different cases

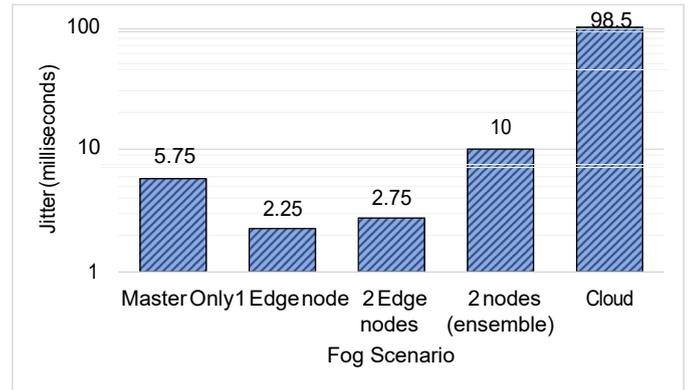

Figure 14: Jitter in different cases

lay. We see that if the task is sent to Broker or any of the edge nodes, then the latency is nearly same as all communication is through single hop data transfers. In ensemble case, the latency is slightly higher. For cloud setting, the latency is very high due to multi-hop transfer of data outside the LAN.

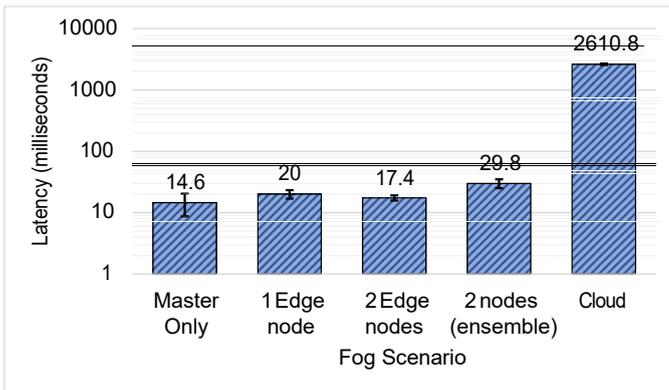

Figure 13: Latency in different cases

Jitter is the variation of response time for consecutive job requests. It is a critical parameter for most real-time applications including health data analysis. Figure 14 (log vertical scale) shows the variation of jitter with the Fog configurations. We observe that jitter is higher for Broker only case compared to the case where tasks are sent to worker nodes. This is because of other tasks including arbitration, resource management and security checking are also performed by Broker. As the workers increase, due to difference in loads of workers jitter slightly increases for two edge nodes compared to single edge node. Jitter is also high in ensemble case. Jitter is very high when tasks are sent to CDC.

Figure 15 shows the variation of execution time. As expected, the execution time in Cloud setup is very low due to higher resource availability. Broker execution time is lesser than the worker nodes as HealthFog workers are Raspberry Pis which have processor with low clock frequency. Also, when ensemble prediction is enabled then the execution time is higher because the worker node now needs to check which class is majority among all predicted classes.

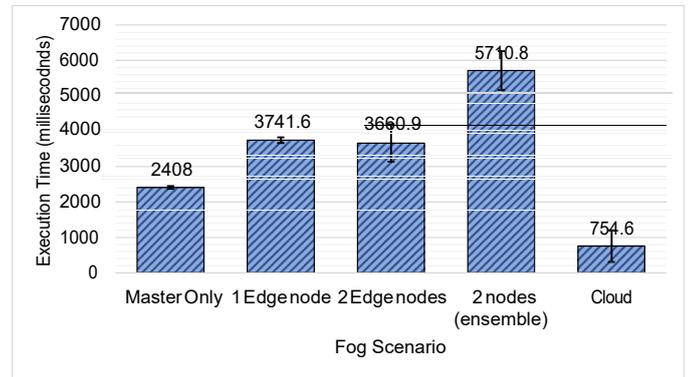

Figure 15: Execution time in different cases

### 7.7. Network Bandwidth Usage Characteristics

Figure 16 shows the variation of Network bandwidth usage of all edge nodes in different scenarios. We see that as the worker nodes increase, the network usage also increase because more heartbeat packets, security checks and data transfer (with cloud) are required. In ensemble case, as data is sent to all worker nodes the network bandwidth consumption is highest.

### 7.8. Power Characteristics

We also tested HealthFog energy consumption characteristics in different scenarios. The power consumption of CDC is very high compared to the Broker node (laptop) or Worker nodes (Raspberry Pi). This leads to very high power consumption in Cloud case compared to Edge case. As the number of Worker nodes increase, the power consumption of the Health-Fog framework also increases.

### 7.9. Analysis with Related Work

Other works that propose computing models for healthcare applications in Fog Computing do not consider various aspects which HealthFog does. Many prior works [13, 16, 17, 22, 23, 39, 42] do not leverage resources close to the edge of the network. As per Figure 13, such models provide a much higher latency as all computation is done on the cloud and hence has higher data transfer times. With the advancement of deep learning based prediction models, HealthFog is able to use state-of-the-art Neural Network models for highly accurate prediction

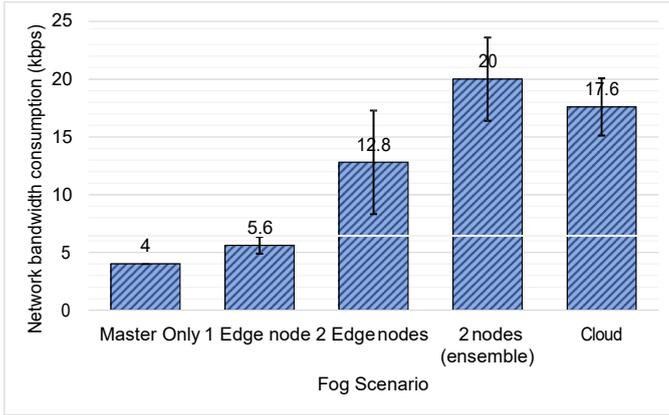

Figure 16: Network usage in different cases

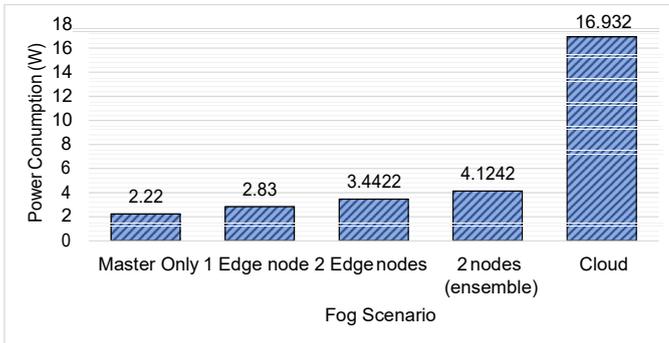

Figure 17: Power consumption in different cases

### 7.10. Discussion and Recommendations

In earlier work [27], the power of FogBus and comparisons with earlier such Fog frameworks were demonstrated showing how FogBus provides more efficient implementation of applications harnessing the Edge and Cloud resources. This work developed a latency and accuracy sensitive application of Heart analysis using the FogBus framework with engineering simplicity and in low time to efficiently use Edge and Cloud resources. The application deployment system provided different configurations that provide better accuracy or latency based on user requirements. Based on the experimental results we propose HealthFog to be used in the following settings based on the target applications:

- For latency critical and lightweight tasks or energy constraint environments, worker nodes should be used. This provides very low result delivery time due to close proximity of worker nodes. If energy and network bandwidth constraints exist then ensemble bagging should be disabled but if not, enabling bagging would give better accuracy.

- For heavy and latency tolerant tasks CDC configuration must be used otherwise such tasks would not be able to successfully complete on resource constraint edge worker nodes.

## 8. Conclusions and Future Work

Healthcare as a service is a huge project. In this research work, we only focus on the healthcare aspects for heart patients by proposing a novel Fog based Smart Healthcare System for Automatic Diagnosis of Heart Diseases using deep learning and IoT called HealthFog. HealthFog provides healthcare as a fog service and efficiently manages the data of heart patients which is coming from different IoT devices. HealthFog integrates deep learning in Edge computing devices and deployed it for a real-life application of Heart Disease analysis. Prior works for such Heart Patient analysis did not utilize deep learning and hence had very low prediction accuracy which renders them useless in practical settings. Deep learning based models with very high accuracy require very high compute resources (CPU and GPU) both for training and prediction. This work allowed complex deep learning networks to be embedded in Edge computing paradigms using novel communication and model distribution techniques like ensembling which allowed high accuracy to be achieved with very low latencies. This was also validated for real-life heart patient data analysis by training neural networks on popular datasets and deploying a working system that provides prediction results in real-time. We used FogBus framework to validate HealthFog in fog computing environment and tested the efficiency of proposed system in terms of power consumption, network bandwidth, latency, jitter, training accuracy, testing accuracy and execution time.

As part of the future work, we propose to extend HealthFog to allow cost-optimal execution given different QoS characteristics and fog-cloud cost models. Currently HealthFog works with file based input data which can be converted to seamlessly of health characteristics of patients. Other works like [2, 46] or [11, 12, 13, 14, 16, 18, 41, 43] lack the ability to integrate such models and hence provide lower disease detection accuracy. This is crucial to provide low latency and highly accurate results in critical healthcare applications especially those concerned with heart related problems like heart attack, stroke or arrhythmia. Furthermore, works that use deep learning [17, 19, 20] do not use ensembling methods to provide even better results by leveraging fog resources for parallel computation and providing significantly higher accuracy. As shown by results in Section 7.4, with ensemble, the prediction accuracy increases by 16% for the case with 5 edge nodes which is significantly higher than what existing systems (not leveraging ensemble deep learning) can provide.

Moreover, unlike prior work HealthFog uses the FogBus framework [27] to provide a diverse set of configurations with different accuracy, response time, network and power usage characteristics. Based on different application and user requirements different configurations can be used as described in the following section. This allows users to customize the framework as per their needs. This non-trivial extension of integration and synchronization among fog computing nodes allows execution ensemble based deep learning models which not only improves disease detection accuracy but is also adaptive as per diverse requirements. Hence, HealthFog provides a novel architecture of healthcare computation not offered by existing works.

integrated to take data directly from sensors to make it user-friendly. Moreover, the model training strategy used currently uses separate training at each worker node. The trained models at each node have combined using various ensemble model of bagging. More intelligent ensemble models can be deployed for further improving the accuracy. Further, proposed architecture can be made robust and generic to incorporate other fog computing applications such as agriculture, healthcare, weather forecasting, traffic management and smart city. HealthFog can also be extended towards other important domains of healthcare such as diabetes, cancer and hepatitis, which can provide efficient services to corresponding patients.

## Software Availability

We released HealthFog as an open source software. The implementation code with experiment scripts and results can be found at the GitHub repository: https://github.com/Cloudslab/HealthFog.

## Acknowledgements

This research work is supported by the Melbourne-Chindia Cloud Computing (MC3) Research Network and Australian Research Council. We would like to thank the editor, area editor and anonymous reviewers for their valuable comments and suggestions to help and improve our research paper. We would also like to thank Samodha Pallewatta, Shashikant Ilager (CLOUDS Lab, University of Melbourne) and Shikhar Tuli (Indian Institute of Technology, Delhi) for their valuable comments on improving the quality of presentation.

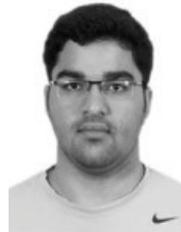

**Shreshth Tuli** is an undergraduate student at the Department of Computer Science and Engineering at Indian Institute of Technology - Delhi, India. He is a national level Kishore Vaigyanic Protsahan Yojana (KVPY) scholarship holder for excellence in science and innovation. He is working as a visiting research fellow at the Cloud Computing and Distributed Systems (CLOUDS) Laboratory, Department of Computing and Information Systems, the University of Melbourne, Australia. Most of his projects are focused on developing technologies for future requiring sophisticated hardware-software integration. His research interests include Internet of Things (IoT), Fog Computing, Network Design, Blockchain and deep learning.

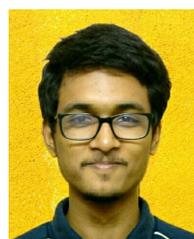

**Nipam Basumatary** is an undergraduate student at Department of Computer Science and Engineering at Indian Institute of Technology, Madras. He is working as a visiting research fellow at Cloud Computing and Distributed Systems (CLOUD) Laboratory, Department of Computing and Information Systems, the University of Melbourne, Australia. His research interests include Machine

Learning, deep learning, Internet of Things (IoT) and Fog Computing.

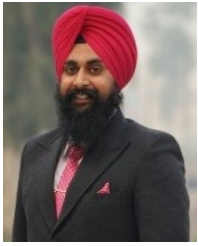

**Sukhpal Singh Gill** is a Lecturer (Assistant Professor) in Cloud Computing at School of Electronic Engineering and Computer Science (EECS), Queen Mary University of London, UK. Prior to this, Dr. Gill has held positions as a Research Associate at the School of Computing and Communications, Lancaster University, UK and also as a Postdoctoral Research Fellow at the Cloud Computing and Distributed Systems (CLOUDS) Laboratory, School of Computing and Information Systems, The University of Melbourne, Australia. Dr. Gill was a research visor at Monash University, University of Manitoba and Imperial College London. He was a recipient of several awards, including the Distinguished Reviewer Award from Software: Practice and Experience (Wiley), 2018, and served as the PC member for venues such as UCC, SE-CLOUD, ICCCN, ICDICT and SCES. His one review paper has been nominated and selected for the ACM 21st annual Best of Computing Notable Books and Articles as one of the notable items published in computing - 2016. He has published more than 50 papers as a leading author in highly ranked journals and conferences with H-index 18. His research interests include Cloud Computing, Fog Computing, Software Engineering, Internet of Things and Big Data. For further information on Dr. Gill, please visit: www.ssgill.me.

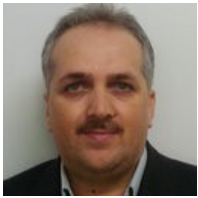

**Mohsen Kahani** is a professor of computer engineering, IT director and head of Web Technology Laboratory at Ferdowsi University of Mashhad and visiting Researcher at Cloud Computing and Distributed Systems (CLOUDS) Laboratory, School of Computing and Information Systems, The University of Melbourne, Australia. His research interests includes semantic web, software engineering, natural language processing and process mining.

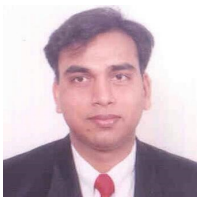

**Rajesh Chand Arya** is a Consultant Cardiac Anesthesiologist at Department of Cardiology, Hero Heart Institute Dayanand Medical College and Hospital, Ludhiana, Punjab, India. He has 12 years of Cardiac Anesthesia experience after post-graduation. Special interest in Trans Esophageal Echo (TEE) & Pediatric, Cardiac Anesthesia. He has delivered more than 20 guest lectures at national & international level.

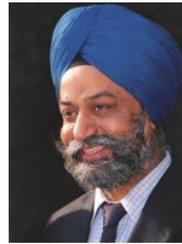

**Gurpreet Singh Wander** is a Chief Cardiologist at Department of Cardiology, Hero Heart Institute Dayanand Medical College and Hospital (DMC & H), Ludhiana, Punjab, India. He joined DMC & H in 1988. He has been awarded BC Roy Award in 2007 and K. Sharan Award by national IMA in 2005. He is a Member of Govern Board of API for 6 years. Dr. Wander has 200+ research publications in top ranked venues which include 5 publications in Nature journals (Nature genetics, Journal of Human Genetics, Scientific Reports and Journal of Human Hypertension).

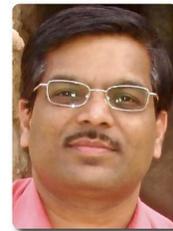

**Rajkumar Buyya** is a Redmond Barry Distinguished Professor and Director of the Cloud Computing and Distributed Systems (CLOUDS) Laboratory at the University of Melbourne, Australia. He is also serving as the founding CEO of Manjrasoft, a spin-off company of the University, commercializing its innovations in Cloud Computing. He served as a Future Fellow of the Australian Research Council during 2012-2016. He has authored over 700 publications and seven text books including "Mastering Cloud Computing" published by McGraw Hill, China Machine Press, and Morgan Kaufmann for Indian, Chinese and international markets respectively. He is one of the highly cited authors in computer science and software engineering worldwide (h-index=127, g-index=281, 84900+ citations). "A Scientometric Analysis of Cloud Computing Literature" by German scientists ranked Dr. Buyya as the World's Top-Cited (1) Author and the World's Most-Productive (1) Author in Cloud Computing. Recently, Dr. Buyya is recognized as a "Web of Science Highly Cited Researcher" in both 2016 and 2017 by Thomson Reuters, a Fellow of IEEE, and Scopus Researcher of the Year 2017 with Excellence in Innovative Research Award by Elsevier for his outstanding contributions to Cloud computing. He served as the founding Editor-in-Chief of the IEEE Transactions on Cloud Computing. He is currently serving as Editor-in-Chief of Journal of Software: Practice and Experience, which was established over 45 years ago. For further information on Dr. Buyya, please visit his cyberhome: www.buyya.com